\documentclass[aps,prl,twocolumn,groupedaddress,nofootinbib,showkeys,nobibnotes]{revtex4-2}

\usepackage{graphicx}
\usepackage{amssymb}
\usepackage{appendix}
\usepackage{hyperref}
\usepackage{enumitem}
\usepackage{amsmath}
\usepackage{physics}

\begin{document}

\title{Liouville's theorem and the foundation of classical mechanics}

\author{Andreas Henriksson}
\email[]{andreas.henriksson@skole.rogfk.no}

\affiliation{Stavanger Katedralskole, Haakon VII's gate 4, 4005 Stavanger, Norway}

%\date{\today}

\begin{abstract}
In this article, it is suggested that a pedagogical point of departure in the teaching of classical mechanics is the Liouville theorem. The theorem is interpreted to define the condition that describe the conservation of information in classical mechanics. The Hamilton equations and the Hamilton principle of least action are derived from the Liouville theorem.

\end{abstract}


\maketitle

\section{1. Introduction}

The theory of classical mechanics is in this article approached from a different perspective. Its purpose is entirely pedagogical. I have taught classical mechanics in the traditional way, starting with Newton's laws of motion, and following up with Hamilton's principle, the Euler-Lagrange equations of motion, the Hamilton equations and the Liouville theorem. The students have, in general, had problems seeing the relations between the standard mathematical representations of classical mechanics. In each class, there are typically a few students that ask whether there exists a foundational principle of classical mechanics that is independent on the specific mathematical representation being chosen. I have never been able to answer this question in a satisfactory manner. This article grew out of the desire to address this question.

Clearly, there is no reason to believe there exist a unique principle. However, in this article a specific point of departure is identified and shown to lead to the traditional formulations. The suggested principle is that of conservation of information. It is argued that the Liouville theorem is the mathematical representation of this principle. The Hamilton equations, the Hamilton principle of least action and the invariance of the Poisson algebra are then understood as different manifestations of the Liouville theorem.

There is nothing new appearing in this article. Everything is known from before. What then, one might ask, is the purpose and use of the article? The answer is threefold. First, and foremost, it suggests an alternative way to teach the subject. As a teacher for many years, it is obvious that it is beneficial to have a diverse repertoar when it comes to presenting and explaining a topic. Personally, I take great pleasure in being able to explain the subject to my students in different ways. Secondly, it provides a different point of view on an old and well-known subject. Eventhough it might not be of any use in the immediate, or near, future, it is generally good to be aware of a multitude of equivalent perspectives on any given problem. Thirdly, to the best of my knowledge, the Hamilton principle has never been derived from the Liouville theorem.

\section{2. Determinism and information}

\begin{figure}[h!]
 \centering
  \includegraphics[width=0.35\textwidth]{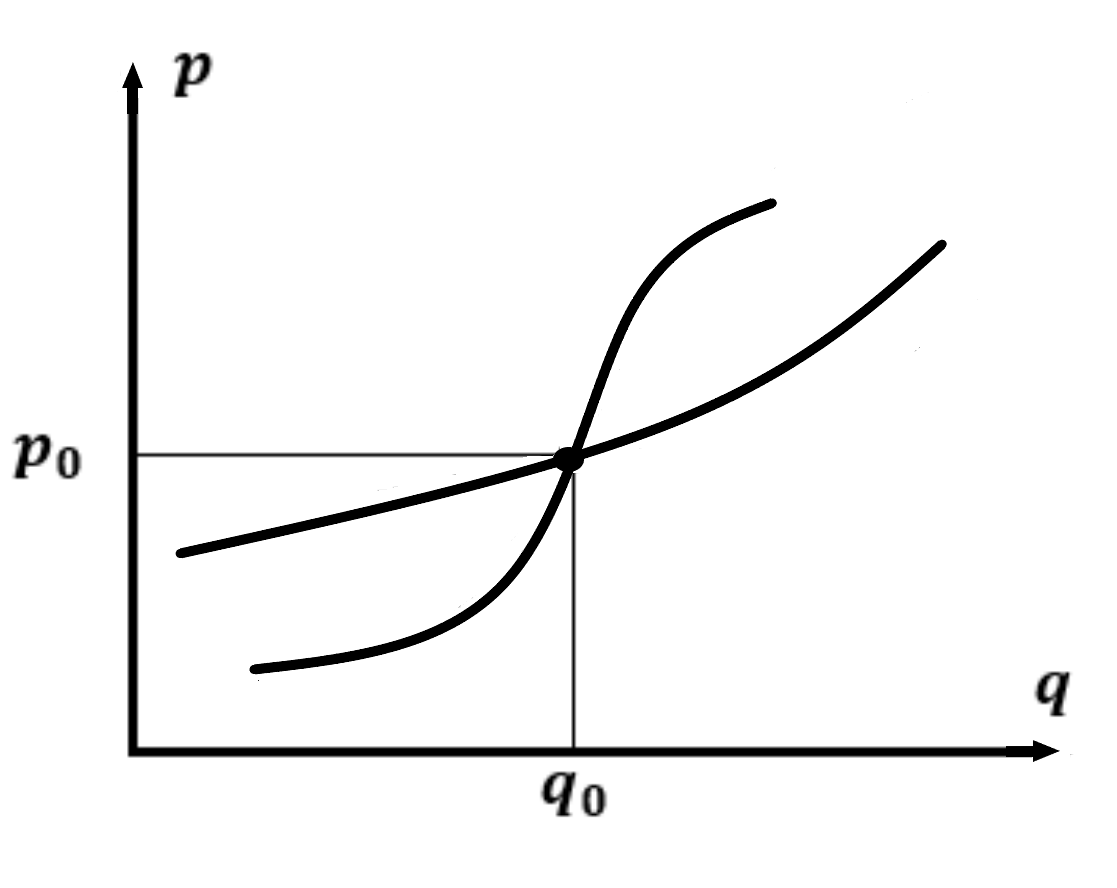}
 \caption{Non-deterministic evolution imply that system trajectories would cross each other on phase space, here at point $(q_0, p_0)$.}
 \label{fig:Fig1}
 \end{figure}

In classical mechanics, it is a fundamental assumption that the evolution of a system is deterministic in both directions of time, i.e. both into the future and into the past. Deterministic evolution of a system mean that it is possible, with absolute certainty, to say that any given state of the system evolved from a definite single state in the past and will evolve into a definite single state in the future. There cannot be any ambiguity in the evolutionary history of a system. Thus, deterministic evolution imply that nowhere on phase space can states converge or diverge, see Fig. \ref{fig:Fig1}.

Systems that appear to evolve non-deterministically give rise to the appearance of irreversible processes in nature. The reason for this is that if a system start out in a given state it is not necessarily the case that the system end up at the same initial state by reversing the motion of the system in time. An example of a seemingly irreversible process is the sliding of a block of cheese along a table. Due to friction the block will always come to rest, apparently independent on the initial condition of the block. Thus, it appear as though the multitude of possible initial states for the block, given by the possibility of sending off the block with different initial speeds, all converge to the same final state where the block is at rest. Knowing the final state of the system does not help in predicting the initial state of the system. Therefore, the experiment with sending off the block of cheese seem to represent an evolution which is non-deterministic into the past.

The origin for the apparent violation of reversibility in physical processes is not due to a fundamental character in physical laws, but rather it is due to the ignorance of the observer. The observer has not taken into account all the details of the system. Degrees of freedom for the system has been ignored. In the case of the sliding block of cheese, it is the individual motion of atoms in the block and table which has been ignored. Assuming that all degrees of freedom for the block and table are followed in perfect detail as the block slide on the table it is clear that each unique initial state will give rise to a unique final state where the distinction between the final states are given by the distinct final position and velocity of each atom in the block and table.

A direct consequence of the assumption of deterministic evolution is that distinctions between physical states never disappear. If there is an initial distinction between states, this distinction will survive throughout the entire motion of the system. That distinctions between states seem to disappear as time unfold is merely a consequence of the difficulty for an observer to keep perfect track of the motion of all particles. In the case of the sliding block, for a human observer, the distinction between individual motions of atoms in the block and table are too small to measure and therefore it appears as though two distinct initial states, characterized by distinct initial speeds, which are easy to measure, converge to the same final state, i.e. that the block is at rest. In conclusion, the assumption of deterministic evolution can equivalently be stated as follows.\\
\\
\textit{The distinction between physical states of a closed system is conserved in time.}\\
\\
Due to the conservation of distinction between physical states, any set of states which lie in the interior of some volume element on phase space will remain interior of this volume element as the system evolve in time.

If a system is followed, as it evolves in time, in perfect detail by an observer, it mean that the observer has perfect and complete knowledge about all the degrees of freedom of the system, i.e. the observer know, with infinite precision, the exact position and momenta of all particles within the system. In such an ideal scenario, the observer has no problem to see the distinction between states of the system. The amount of knowledge, or information, about the system possessed by the observer, at any instant of time, is complete. Since the ideal observer never lose track of the system, the distinction between states is never lost. In other words, the knowledge, or information, that the observer has about the system is not lost as the system evolve in time.

If, however, as is the case in practical reality, the observer has a limited ability to track the motion of individual particles, the observer do not possess complete information about the system. Even worse, the observer may, as is usually the case for complicated systems with many degrees of freedom, find it more and more difficult to track the system as time unfold. In such a scenario, the amount of information about the system, possessed by the observer, decrease with time. In other words, from the perspective of the ignorant observer, information about the system is lost. However, it is important to emphasize that this apparent loss of information is entirely due to the ignorance of the observer. If all the degrees of freedom were tracked with infinite precision, information would never be lost. In the case of the sliding block of cheese, the observer has lost information because the system was known to exist in one of two distinct initial states, obtained by measuring the initial speed of the block, whereas it is not possible to distinguish between the two final states.

In conclusion, the loss of distinction between states imply that information has been lost. Thus, the conservation of distinction between states can equivalently be stated as an assumption of information conservation:\\
\\
\textit{The information contained within a closed system is conserved in time.}\\
\\
In other words, the assumption that classical systems evolve deterministically, i.e. that the state of the system is perfectly predictable by an observer both into the future and back to the past, is equivalent to the statement that an observer of the system possess complete information about the system, and assuming that the system is closed, this amount of information is never lost.

\section{3. The Liouville theorem}

Consider an arbitrary region $\Omega$ on a $2$-dimensional phase space, with volume $V_\Omega$ and volume element $\Delta q\Delta p$. The mathematical condition imposing information conservation is
\begin{equation}
\label{eq:numberconstant}
\frac{\Delta N}{\Delta t}=0,
\end{equation}
where $N$ is the number of states within the phase space volume $\Omega$. The condition state that $N$ can neither increase nor decrease within the time interval $\Delta t$. For this condition to be satisfied, it is necessary that the incoming and outgoing flow of states through $\Omega$ within $\Delta t$ cancel, i.e. that
\begin{equation}
\Delta\left(\rho(q,p)\dot{q}\right)+\Delta\left(\rho(q,p)\dot{p}\right)=0,
\end{equation}
where $\rho(q,p)$ is the density of states on phase space, and the flow differences are defined by, respectively,
\begin{equation}
\Delta\left(\rho(q,p) \dot{q}\right)\equiv \left\{\rho(q_{out},p)\dot{q}_{out}-\rho(q_{in},p)\dot{q}_{in}\right\} \Delta p
\end{equation}
and
\begin{equation}
\Delta\left(\rho(q,p) \dot{p}\right)\equiv \left\{\rho(q,p_{out})\dot{p}_{out}-\rho(q,p_{in})\dot{p}_{in}\right\} \Delta q.
\end{equation}
In differential form the condition, after having been extended to be valid for an arbitrary length of time, read in vector notation as
\begin{equation}
\label{eq:continuityequationhamiltonianflow}
\frac{\partial\rho}{\partial t}+\vb{\nabla}\cdot\left(\rho\vb{v}\right)=0,
\end{equation}
where 
\begin{equation}
\vb{\nabla}\equiv\left(\frac{\partial}{\partial q}, \frac{\partial}{\partial p}\right)
\end{equation}
is the differential operator on phase space, and
\begin{equation}
\vb{v}\equiv\left(\dot{q}, \dot{p}\right)
\end{equation}
is the velocity by which states flow on phase space. Equation \ref{eq:continuityequationhamiltonianflow} is the Liouville continuity equation \cite{bloch} for the density of states on phase space. It say that the number of states is locally conserved. The term $\vb{\nabla}\cdot\left(\rho \vb{v}\right)$ represent the net flow of states through $\Omega$, i.e. the difference between the outflow and inflow of states. The continuity equation can be rewritten as
\begin{equation}
\frac{d\rho}{dt}+\rho\ \vb{\nabla}\cdot\vb{v}=0
\end{equation}
by using the total time derivative of the density of states and the product rule applied to the net flow of states. Thus, if the divergence of the phase flow velocity vanishes, i.e. if
\begin{equation}
\vb{\nabla}\cdot\vb{v}=0,
\end{equation}
then, by the continuity equation, the density of states on phase space is constant in time along the flow on phase space, i.e.
\begin{equation}
\frac{d\rho}{dt}=0.
\end{equation}
In such a situation, the flow of the system on phase space is incompressible because the condition that the density of states at any given location $(q,p)$ on phase space, within an arbitrary region $\Omega$, do not change over time ensure that the states do not lump together. In other words, in conclusion, a necessary and sufficient condition for the flow of the system on phase space to conserve information is that the divergence of the phase flow velocity vanish. This is the Liouville theorem\cite{bloch}\footnote{To the best of the authors knowledge, the physical formulation and relevance of the Liouville theorem was first stated by J.W. Gibbs in 1902\cite{gibbs}. There it was referred to as the "Principle of conservation of density-in-phase" or equivalently as the "Principle of conservation of extension-in-phase". However, the mathematical background for the theorem dates back to J. Liouville in 1838\cite{liouville1}.}.

The $2$-dimensional Liouville theorem straightforwardly generalize to $6N$-dimensional phase space. Each conjugate pair $(q_j,p_j)$, where $j\in [1, 3N]$, give rise to an independent Liouville continuity equation, i.e.
\begin{equation}
\frac{d\rho_j}{dt}+\rho_j\ \vb{\nabla}\cdot\vb{v_j}=0, \ j\in[1, 3N],
\end{equation}
where $\rho_j\equiv \rho(q_j, p_j)$ is the density of states in the $2$-dimensional subset $(q_j, p_j)$ of the $6N$-dimensional phase space and $\vec{v}_j\equiv (\dot{q}_j, \dot{p}_j)$ is the phase flow velocity along this subset. Thus, information is conserved on the $6N$-dimensional phase space if the divergence of each phase flow velocity $\vec{v_j}$ vanish, i.e. if
\begin{equation}
\vb{\nabla}\cdot\vb{v_j}=0 \ \ \forall j\in[1, 3N].
\end{equation}

\section{4. Hamilton's equations}

The vanishing divergence of the flow velocity $\vb{v_j}$ for all conjugate pairs $(q_j, p_j), j\in[1, 3N]$, written out explicitly in terms of its velocity components $\dot{q}_j$ and $\dot{p}_j$, become
\begin{equation}
\frac{\partial\dot{q}_j}{\partial q_j}+\frac{\partial\dot{p}_j}{\partial p_j}=0\ \ \forall j\in[1, 3N].
\end{equation}
Let $\mathcal{H}$ be a smooth function on the $6N$-dimensional phase space with the property that it contains no terms that mix different conjugate pairs, e.g. $p_i\cdot p_j, \ \forall i\neq j$. In this situation, taking into account that the set of conjugate pairs $\left\{(q_j, p_j)\right\}_{j=1}^{3N}$ are postulated to be independent, the condition of vanishing divergence can equivalently be stated by the set of differential equations known as Hamilton's equations,
\begin{eqnarray}
\label{eq:hamiltonsequations1}
\dot{q}_j&=&\frac{\partial \mathcal{H}}{\partial p_j}\ \ \forall j\in[1, 3N],\\
\dot{p}_j&=&-\frac{\partial \mathcal{H}}{\partial q_j}\ \ \forall j\in[1, 3N].
\label{eq:hamiltonsequations2}
\end{eqnarray}
Under these circumstances, the Hamilton equations are, according to the Liouville-Arnold theorem \cite{liouville2}\cite{arnold}, integrable from a set of known initial conditions. This simply mean that they describe an evolution of the system which is unique and deterministic. Thus, given the function $\mathcal{H}$, the flow of the system in time is determined by how $\mathcal{H}$ change on phase space. In this sense, $\mathcal{H}$ is said to be the generator for the motion in time of the system on phase space. The flow of the system on phase space, described by the Hamilton equations, is referred to as an Hamiltonian flow.

\section{5. The Hamiltonian and Lagrangian}

\begin{figure}[h!]
 \centering
  \includegraphics[width=0.35\textwidth]{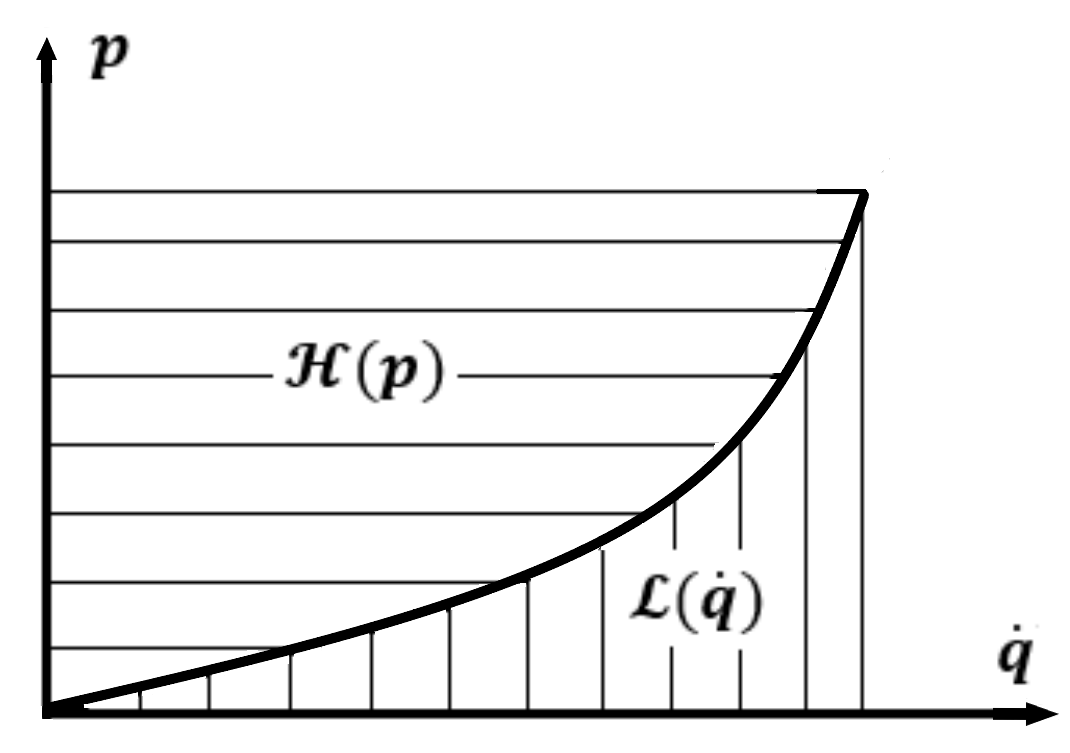}
 \caption{The areas under $\dot{q_j}(p_j)$ and $\dot{p_j}(q_j)$ graphs define the Hamiltonian and Lagrangian, respectively.}
 \label{fig:Fig2}
 \end{figure}

Equation \ref{eq:hamiltonsequations1}, for a specific conjugate pair $(q_j, p_j)$, correspond to the integral equation
\begin{equation}
\mathcal{H}(p_j)=\int dp_j\ \dot{q_j}(p_j).
\end{equation}
The momentum $p_j$  and speed $\dot{q_j}$ are assumed to be in one-to-one correspondence. This mean that for each value of $\dot{q_j}$ there is a unique value for $p_j$, and vice versa. The function $\mathcal{H}(p_j)$ is then geometrically interpreted as the unique area under the $\dot{q_j}(p_j)-$graph, bounded by $(0, p_j)$ and $(0, \dot{q_j}(p_j))$, see Fig. \ref{fig:Fig2}.
Due to the one-to-one correspondence between $p_j$ and $\dot{q_j}$ it is possible to define a related area, $\mathcal{L}(\dot{q_j})$, given by the unique area under the $p_j(\dot{q_j})-$graph,
\begin{equation}
\mathcal{L}(\dot{q_j})=\int d\dot{q_j}\ p_j(\dot{q_j}).
\end{equation}
This integral equation corresponds to the differential equation
\begin{equation}
\frac{d\mathcal{L}(\dot{q_j})}{d\dot{q_j}}=p_j.
\end{equation}
The total area of the rectangle bounded by $(0,p_j)$ and $(0,\dot{q_j})$ is given by
\begin{equation}
\mathcal{L}(\dot{q_j})+\mathcal{H}(p_j)=p_j\cdot \dot{q_j}.
\end{equation}
It is possible to include a dependence on the generalized coordinate $q_j$ under the constraint that any $q_j-$dependent terms in the functions $\mathcal{H}$ and $\mathcal{L}$ cancel such that the total area is $q_j-$independent. Thus, in general, the functions $\mathcal{H}$ and $\mathcal{L}$, referred to as the Hamiltonian and Lagrangian, respectively, satisfy the so-called Legendre transformation, i.e.
\begin{equation}
\mathcal{L}(q_j, \dot{q_j})+\mathcal{H}(q_j, p_j)=p_j\cdot \dot{q_j},
\end{equation}
where
\begin{eqnarray}
\mathcal{L}(q_j,\dot{q_j})&=&\int_0^{\dot{q_j}}\ d\dot{q_j}\ p_j(\dot{q_j})-U(q_j),\\
\mathcal{H}(q_j, p_j)&=&\int_0^{p_j}\ dp_j\ \dot{q_j}(p_j)+U(q_j).
\end{eqnarray}
The requirement that the total area is $q_j-$independent cause the Hamiltonian and Lagrangian to have a relative sign difference for the function $U(q_j)$.

For the $6N$-dimensional phase space, the Hamiltonian and Lagrangian are defined by
\begin{eqnarray}
\mathcal{L}(q,\dot{q})&\equiv&\sum_{j=1}^{3N}\int_0^{\dot{q_j}}\ d\dot{q_j}\ p_j(\dot{q_j})-U(q).\\
\mathcal{H}(q, p)&\equiv&\sum_{j=1}^{3N}\int_0^{p_j}\ dp_j\ \dot{q_j}(p_j)+U(q),\\
\end{eqnarray}
where the function $U(q)$, defined by
\begin{equation}
U(q)\equiv \sum_{j=1}^{3N}U(q_j),
\end{equation}
is referred to as the potential energy of the system.

\section{6. Principle of stationary action}

The pair of Hamilton equations
\begin{eqnarray}
-\frac{\partial\mathcal{H}}{\partial q_j}-\dot{p_j}&=&0,\\
\dot{q_j}-\frac{\partial\mathcal{H}}{\partial p_j}&=&0,
\end{eqnarray}
is the local, differential, representation of the principle of information conservation on phase space. A global, or integral, representation can be obtained by considering the entire evolutionary path from some initial time $t_i$ to some final time $t_f$ where the Hamilton equations are integrated over time\footnote{For the derivation of an integral representation on configuration space starting from Newton's second law of  motion, see chapter 10 in reference \cite{jeffreys}.}. For this purpose, multiply the Hamilton equations with two independent arbitrary functions of time, $\delta q_j(t)$ and $\delta p_j(t)$, representing, respectively, small displacements in $q_j$ and $p_j$ on phase space, in the following manner,
\begin{eqnarray}
\label{eq:hd1}
\left(-\frac{\partial\mathcal{H}}{\partial q_j}-\dot{p_j}\right)\delta q_j(t)&=&0,\\
\label{eq:hd2}
\left(\dot{q_j}-\frac{\partial\mathcal{H}}{\partial p_j}\right)\delta p_j(t)&=&0.
\end{eqnarray}
The displacements $\delta q_j(t)$ and $\delta p_j(t)$ are pictured as slight variations of the physical path on phase space, i.e.
\begin{eqnarray}
q_j(t)&\rightarrow & q_j(t)+\delta q_j(t),\\
p_j(t)&\rightarrow & p_j(t)+\delta p_j(t).
\end{eqnarray}
Equations \ref{eq:hd1} and \ref{eq:hd2} are equivalent to the Hamilton equations since they hold for arbitrary variations. The fact that it is necessary to introduce two displacement functions is due to the independence of the state parameters $q_j$ and $p_j$. The boundary conditions are given by 
\begin{eqnarray}
\delta q_j(t_i)&=&\delta q_j(t_f)=0,\\
\delta p_j(t_i)&=&\delta p_j(t_f)=0,
\end{eqnarray}
i.e. the variations vanish at the initial and final times. Integrating the Hamilton equations over time from $t_i$ to $t_f$ give, to leading order in the variations,
\begin{equation}
\int_{t_i}^{t_f}\ dt\left[\left(-\frac{\partial\mathcal{H}}{\partial q_j}-\dot{p_j}\right)\delta q_j(t)+\left(\dot{q_j}-\frac{\partial\mathcal{H}}{\partial p_j}\right)\delta p_j(t)\right]=0.
\end{equation}
Integration by parts and recalling the boundary conditions gives
\begin{equation}
\delta\mathcal{A}(q_j, \dot{q_j})=0,
\end{equation}
where
\begin{equation}
\mathcal{A}(q_j, \dot{q_j})\equiv \int_{t_i}^{t_f}\ dt\ \mathcal{L}(q_j, \dot{q_j})
\end{equation}
is the action of the system within the subset $(q_j, p_j)$ on the $6N-$dimensional phase space. The action on the entire phase space is given by
\begin{eqnarray}
\mathcal{A}(q,\dot{q})&\equiv &\sum_{j=1}^{3N} \int_{t_i}^{t_f}\ dt\ \mathcal{L}(q_j, \dot{q_j})\nonumber\\
&=& \int_{t_i}^{t_f}\ dt\ \mathcal{L}(q, \dot{q}).
\end{eqnarray}
This is Hamilton's formulation of the principle of stationary action, or shortly, Hamilton's principle. It is a global representation of information conservation, i.e. a statement on the entire evolutionary path which must be satisfied if the system is to adhere to the principle of information conservation.

Since the Hamilton principle can be derived from the Hamilton equations, which in turn is an immediate consequence of the requirement that the divergence of the Hamiltonian flow velocity vanish, it should be possible to obtain the Hamilton principle directly from the requirement that $\vb{\nabla}\cdot\vb{v_j}=0$ is invariant under the displacements $\delta q_j(t)$ and $\delta p_j(t)$. Given that the variations are small, the flow velocity $\vec{v_j}$ can be expanded as a Taylor series about the state $(q_j, p_j)$ where terms that are of quadratic, or higher, order in the variations $\delta q_j$ and $\delta p_j$ can be ignored. The infinitesimal change in $\vb{v_j}$ thus become
\begin{eqnarray}
\delta\vb{v_j}&=&\vb{v_j}(q_j+\delta q_j, p_j+\delta p_j)-\vb{v_j}(q_j,p_j)\nonumber\\
&=&\delta q_j \frac{\partial}{\partial q_j} \vb{v_j}+\delta p_j \frac{\partial}{\partial p_j} \vb{v_j}.
\end{eqnarray}
The divergence of the flow velocity transform as
\begin{equation}
\vb{\nabla}\cdot\vb{v_j}\rightarrow \vb{\nabla}\cdot\left(\vb{v_j}+\delta\vb{v_j}\right)=\vb{\nabla}\cdot\vb{v_j}+\vb{\nabla}.\cdot\delta\vb{v_j}
\end{equation}
If $\vb{\nabla}\cdot\delta\vb{v_j}\neq 0$, information is not conserved for the deviated path. Therefore, it is required that
\begin{equation}
\vb{\nabla}\cdot\delta\vb{v_j}=0,
\end{equation}
which is equivalent to
\begin{equation}
\delta\left(\vb{\nabla}\cdot\vb{v_j}\right)=0.
\end{equation}
This statement is for a blob of volume $dV$ which enclose the single state $(q_j, p_j)$. Information conservation should hold for all varied states along the evolutionary path of the system, from the initial state $(q_j, p_j)_i$, at time $t_i$, to the final state $(q_j, p_j)_f$, at time $t_f$. Thus, the above statement should be integrated over all blobs of volume $dV$ along the path, i.e. the integration is over a tube, with volume $V$, whose interior define the region of extended phase space where the principle of information conservation is fulfilled. Thus,
\begin{equation}
\delta \int_{t_i}^{t_f}dt\int_{V}dV\ \vb{\nabla}\cdot\vb{v_j} =0.
\end{equation}
Applying the divergence theorem
\begin{equation}
\int_{V}dV\  \vb{\nabla}\cdot\vb{v_j}=\int_{\partial V} \vb{dS}\cdot\vb{v_j},
\end{equation}
gives
\begin{equation}
\delta \int_{t_i}^{t_f}dt\int_{\partial V} \vec{dS}\cdot\vb{v_j}=0.
\end{equation}
The integrand $\vb{dS}\cdot\vb{v_j}$ represent the density of the net Hamiltonian flow out of the tube. The surface area element $\vb{dS}$ is given by
\begin{equation}
\vb{dS}=dS\ \vb{n},
\end{equation}
where $\vb{n}=\left(p_j, q_j\right)$ is the normal vector to the surface of the tube, i.e. $\vb{n}$ give the direction in phase space in which the system has to flow if it is to eventually reach a region where the principle of conservation of information no longer hold. Thus, with $\vb{v_j}=\left(\dot{q_j}, \dot{p_j}\right)$, the integrand becomes
\begin{equation}
\left(p_j, q_j\right) \cdot \left(\dot{q_j}, \dot{p_j}\right)=p_j\dot{q_j}+q_j\dot{p_j}.
\end{equation}
Using that $q_j=\int dq_j$ and the Hamilton equation $\dot{p_j}=-\frac{\partial\mathcal{H}}{\partial q_j}$, the integrand can be written as
\begin{equation}
p_j\dot{q_j}-\int dq_j \frac{\partial\mathcal{H}}{\partial q_j}=p_j\dot{q_j}-\int d\mathcal{H}=p_j\dot{q_j}-\mathcal{H}.
\end{equation}
Equivalently, the integrand could have been written as
\begin{equation}
q_j\dot{p_j}+\mathcal{H},
\end{equation}
by using that $p_j=\int dp_j$ and the other Hamilton equation $\dot{q_j}=\frac{\partial\mathcal{H}}{\partial p_j}$. However, the form $p_j\dot{q_j}-\mathcal{H}$ is the preferred choice due to the fact that it is equal to the Lagrangian $\mathcal{L}(q_j, \dot{q_j})$. Thus, on the $6N$-dimensional phase space it is obtained that
\begin{equation}
\delta \int_{t_i}^{t_f}dt\int dS \ \mathcal{L}=0.
\end{equation}
The equality must hold independently on the surface area of the tube, i.e. the principle of information conservation should hold true independently on the number of states in which the system can exist. Therefore, the integration over the surface area can be taken outside of the infinitesimal variation, giving that
\begin{equation}
\delta \int_{t_i}^{t_f}dt \ \mathcal{L}=0,
\end{equation}
which is, again, Hamilton's principle. Thus, the Hamilton principle can be derived directly from the Liouville theorem.

\section{7. Invariance of the Poisson algebra}

Given that the divergence of the Hamiltonian flow velocity vanish, the Liouville equation can be written as
\begin{equation}
\frac{\partial\rho}{\partial t}+\vb{\nabla}\rho\cdot\vb{v}=0.
\end{equation}
The Poisson bracket $\left\{\rho, \mathcal{H}\right\}$ between the density of states $\rho$ and the Hamiltonian $\mathcal{H}$ is defined by
\begin{equation}
\left\{\rho, \mathcal{H}\right\}\equiv \vb{\nabla}\rho\cdot\vb{v}=\frac{\partial\rho}{\partial q}\frac{\partial\mathcal{H}}{\partial p}-\frac{\partial\rho}{\partial p}\frac{\partial\mathcal{H}}{\partial q}.
\end{equation}
In general, the Poisson bracket $\left\{A, B\right\}$ between any two arbitrary functions $A$ and $B$ on phase space is defined by
\begin{equation}
\left\{A, B\right\}\equiv \frac{\partial A}{\partial q}\frac{\partial B}{\partial p}-\frac{\partial A}{\partial p}\frac{\partial B}{\partial q}.
\label{eq:poissonbracket}
\end{equation}
In this notation, the Hamilton's equations are written as
\begin{eqnarray}
\dot{q}&=&\left\{q, \mathcal{H}\right\},\\
\dot{p}&=&\left\{p, \mathcal{H}\right\}.
\end{eqnarray}
The Poisson bracket satisfy a set of algebraic properties. It is antisymmetric, i.e.
\begin{equation}
\left\{A, B\right\}=-\left\{B, A\right\}.
\end{equation}
It satisfy linearity, i.e.
\begin{equation}
\left\{a A+ b B, C\right\}=a\left\{A, C\right\}+b\left\{B, C\right\}.
\label{eq:poissonlinearity}
\end{equation}
Furthermore, it satisfy the product rule and the Jacobi identity, i.e.
\begin{equation}
\left\{AB, C\right\}=A\left\{B, C\right\}+\left\{A, C\right\}B,
\label{eq:poissonproduct}
\end{equation}
\begin{equation}
\left\{A, \left\{B, C\right\}\right\}+\left\{B, \left\{C, A\right\}\right\}+\left\{C, \left\{A, B\right\}\right\}=0.
\end{equation}
These properties define the Poisson algebra of classical mechanics. Since the Liouville equation for the incompressible Hamiltonian flow can be expressed in terms of the Poisson bracket, the Liouville theorem can equivalently be stated by saying that the evolution in time of any given system conserve information if it leave the Poisson algebra invariant.

\section{8. Conclusion}

The Liouville theorem is interpreted as the mathematical condition representing the physical conservation of information in classical mechanics. The Hamilton equations, the Hamilton principle and the invariance of the Poisson algebra are distinct, but equivalent, manifestations of the theorem.

\end{document}